\documentclass[published]{JHEP3} 
\JHEP{00(2009)000}
\JHEPspecialurl{http://jhep.sissa.it/JOURNAL/JHEP3.tar.gz}
\usepackage{epsfig,multicol,bbm}
\usepackage{dcolumn}
\usepackage{bm}
\usepackage{latexsym}
\usepackage{bm}
\usepackage{pifont}
\usepackage{amsmath,amsfonts,latexsym,amstext,amsbsy,amsthm}
\usepackage{amssymb}
\usepackage{setspace}

\newcommand\fverb{\setbox\fverbbox=\hbox\bgroup\verb}
\newcommand\fverbdo{\egroup\medskip\noindent%
			\fbox{\unhbox\fverbbox}\ }
\newcommand\fverbit{\egroup\item[\fbox{\unhbox\fverbbox}]}
\newbox\fverbbox

\title{Origin of $10^{15}-10^{16}$ G  Magnetic Fields in  the Central Engine of Gamma Ray Bursts}

\author{Rafael S. de Souza\\IAG, Universidade de S\~{a}o Paulo, Rua do Mat\~{a}o 1226, Cidade
Universit\'{a}ria, \\ CEP 05508-900, S\~{a}o Paulo, SP, Brazil\\
E-mail: \email{rafael@astro.iag.usp.br}}
\author{Reuven Opher\\IAG, Universidade de S\~{a}o Paulo, Rua do Mat\~{a}o 1226, Cidade
Universit\'{a}ria, \\ CEP 05508-900, S\~{a}o Paulo, SP, Brazil\\
E-mail: \email{opher@astro.iag.usp.br}}

\received{\today} 		
\accepted{\today}

\abstract{Various authors have suggested that the gamma-ray burst (GRB) central engine is a rapidly rotating,  strongly magnetized,  $(\sim 10^{15}-10^{16}$ G) compact object. The strong magnetic field can accelerate and collimate the relativistic flow and the rotation of the compact object can be the energy source of the GRB. The major problem in this scenario is the difficulty of finding an astrophysical mechanism for obtaining such intense fields. Whereas, in principle, a neutron star could maintain such strong fields, it is difficult to justify a scenario for their creation. If the compact object is a black hole,  the problem is more difficult since,  according to general relativity it has ``no hair" (i.e., no magnetic field).   Schuster, Blackett, Pauli,  and others have suggested that a rotating neutral body can create a magnetic field by non-minimal gravitational-electromagnetic coupling (NMGEC). The Schuster-Blackett form of NMGEC was obtained from the Mikhail and Wanas's  tetrad theory of gravitation (MW). We call the general theory NMGEC-MW. 
 We investigate here
 the possible origin of the intense magnetic fields $\sim 10^{15}-10^{16}$  G in GRBs
 by
  NMGEC-MW.
   Whereas these  fields
   are difficult to explain astrophysically, we find that they are easily explained by NMGEC-MW. It not only explains the origin of the $\sim 10^{15}-10^{16}$G fields when the compact object is a neutron star, but also when it is a black hole.}


\begin{document}

\section{Introduction}

Gamma-ray bursts (GRBs) are short and intense pulses of soft $\gamma$ rays. The bursts last from a fraction of a second to several hundred seconds.    
Magnetic fields in GRBs  are believed to play  a  crucial role in the internal engine as a possible way to power and collimate  the relativistic outflow   \cite{pir05b,lee00}. Energy considerations  require  extremely large magnetic fields on the order of $10^{15}$G  \cite{pir05,uso92,lee00}. 
 The relativistic outflow is a Poynting flux (with negligible baryon
 content) \cite{pir05}. 

 Usov \cite{uso92} suggested that GRBs arise during the formation of rapidly rotating  highly magnetized neutron stars.  Very strong magnetic fields could  form and the pulsar could produce a relativistic Poynting flux flow.  Various authors have investigated this magnetar-GRB connection \cite{tho94, kat97, klu98, whe00, tho04, met07, buc06, kom07, buc07, yu07}.

Thompson \cite{tho07} reviewed the arguments for proto-magnetars as possible GRB engines. Long duration GRBs ($\sim$ several hundred seconds) are, in general, attributed to core-collapse supernova. A core-collapse supernova leaves behind a hot proto-neutron star that cools on the Kelvin-Helmholtz time scale $\tau_{KH}~ (\sim 10-100 s$) by radiating neutrinos. Assuming that neutrino absorption on free nucleons dominates heating, a characteristic thermal pressure is $\sim 3 \times 10^{28} ergs ~cm^{-3}$. Setting $B^{2}/8\pi$ equal to this pressure requires $B\sim 10^{15}$ G. Thus if $B$ is involved in collimating and accelerating the relativistic jet, a field $B\gtrsim 10^{15}$ G is necessary. Like beads on a wire, the magnetic field lines force the plasma around the proto-magnetar into co-rotation with the stellar surface out to the Alfven surface, where the magnetic energy density equals the kinetic energy density of the outflow. A neutron star with a millisecond spin period has a reservoir   of rotational energy of $\sim 2 \times 10^{52} P^{-2} ergs$, where $P$ is the period of rotation of the neutron star in milliseconds. Thus, the rotational energy of the proto-magnetar   can be tapped and could produce a GRB if its rotational period is $\sim$ milliseconds. 

In summary, the  reasons for considering millisecond proto-magnetars as the central engine of long-duration GRB are:
\begin{enumerate}
\item the reservoir of rotational energy is in the range required to power GRBs;
\item the characteristic time of the long-duration GRBs is on the order  of the Kelvin-Helmholtz cooling time, $\tau_{KH}$;
\item there is an observational connection between core collapse supernovae and long duration GRBs; and
\item the magnetic field  of a magnetar $\gtrsim 10^{15}$ G is strong enough to accelerate and collimate the outflow. 
\end{enumerate}

The  major problem of  the magnetar model is the origin of the $\sim 10^{15}-10^{16}$ G fields. As noted by Spruit \cite{spr08}, inheritance of the magnetic field from a main sequence progenitor and dynamo action at some stage of the progenitor is not sufficient to explain the fields of magnetars. A runaway phase of exponential growth would be needed to achieve sufficient field amplification. The  magnetorotational instability is a possibility, but the magnetic  field would probably decay rapidly by magnetic instabilities.

  In a core collapse supernova, a black hole can be formed instead of a neutron star. The justification of the presence of strong magnetic fields in that  case is more difficult since, according to general relativity,  black holes have no magnetic fields.
  
  The Schuster-Blackett (S-B) conjecture \cite{sch11, bla47}  suggests a
gravitational origin of the magnetic fields in rotating neutral
bodies, i.e., a  non-minimal gravitational-electromagnetic coupling (NMGEC).  An early attempt to make a theory that encompasses  the S-B conjecture  was made  by  Pauli  \cite{pau33}.
Latter, attempts were made by   Bennet \textit{et al.} \cite{ben49},
Papapetrou \cite{pap50} , Luchack  \cite{luc52},  and Barut \cite{bar85}. 
 The majority of these studies were based on the five-dimensional
Kaluza-Klein formalism. This formalism was used in order  to
describe a unified theory of gravitation and electromagnetism  in such a way so  that the S-B conjecture is obtained.
Opher and Wichoski \cite{oph97} applied the S-B conjecture for the   origin of magnetic fields in rotating galaxies and proposed that the $B \sim 10^{-6}-10^{-5}$ G magnetic
fields in spiral galaxies are directly obtained from NMGEC.
  Mikhail \textit{et al.} \cite{ mik95} showed that Mikhail and Wanas tetrad theory of gravitation (MW) \cite{mik77} predicts the S-B conjecture of NMGEC. We call this the NMGEC-MW theory.
  
  We investigate here  the possibility that NMGEC-MW is the origin
of the intense magnetic fields $10^{15}-10^{16}$ G, 
 connected with  GRBs produced by rotating neutron stars or black holes. In section 2,  we discuss NMGEC and in section 3,  NMGEC-MW. In section 4,  we apply NMGEC-MW to GRBs. Our conclusions and discussion are presented in section 5.

\section{NMGEC}

The Schuster-Blackett conjecture relates  the angular momentum
\textbf{L} to the magnetic dipole moment \textbf{m}:
\begin{equation}
\textbf{m}=\left[\beta \frac{\sqrt{G_{\rm N}}}{2c}\right]\textbf{L},
\label{m}
\end{equation}
where $\beta$ is approximately a constant,  on the order of unity \cite{oph97}, $G_{\rm N}$  the Newtonian constant of
gravitation, and c is  the speed of light.  The angular momentum
\textbf{L} is

\begin{equation}
\mathbf{L}= I \mathbf{ \Omega},
\end {equation}
where $\mathbf{ \Omega} = 2 \pi P^{-1}$ is the angular velocity, P
the rotational period, and I is the moment of inertia. The
 dipole moment $\mathbf{m}$ is related to the magnetic field $\mathbf{B}$ by

\begin{equation}
\mathbf{B} = \frac{3(\mathbf{m}\cdot
\mathbf{r})\mathbf{r}-\mathbf{m}\mathbf{|r|}^{2}}{\mathbf{|r|}^{5}},
\end{equation}
where $\mathbf{r}$ is the distance from $\mathbf{m}$ to the point at which $\mathbf{B}$ is measured.

The observations and experiments
 supporting the S-B conjecture include
the early work of  Blackett  \cite{bla47} and  Wilson \cite{wil23}.   Observational evidence for
the S-B conjecture was compiled by Sirag \cite{sir79},  who compared
the predictions of Eq.(\ref{m}) to the observed values of the ratio
of the magnetic moment to the angular momentum for the Earth,
Sun, the star 78 Vir, the Moon, Mercury, Venus, Jupiter, Saturn, and
the neutron star Her X-1.   The value of $\beta$ for all of  these objects was $\sim 0.1$ to within a factor of ten.

\section{NMGEC-MW}

A possibility for obtaining the Schuster-Blackett relation from first principles was proposed by Mikhail \textit{et al.} \cite{mik95}. They  used one of the solution of the generalized tetrad field theory of Mikhail and Wanas \cite{mik77}. The solution used is of the type FIGI \cite{mik81} which can represent a generalized non-zero electromagnetic and gravitational field outside a neutral rotating spherically symmetric body \cite{mik85}. Two main reasons why MW used the tetrad space for the formalism of the generalized field theory are:
\begin{enumerate}
\item It is based on 16 independent field variables instead of the 10 variables $g_{\mu\nu}$ of the ordinary Riemannian space; and
\item It has 11 independent tensors in space of the two symmetric tensors $g_{\mu\nu}$ and $R_{\mu\nu}$ of ordinary Riemannian space.

\end{enumerate}

 Mikhail \textit{et al. } \cite{mik95} evaluated the field equations,  assuming a symmetric solution  for the Reissner-Nordstr\"{o}m metric,
\begin{eqnarray}
&&ds^2=g_{\mu \nu }dx^\mu dx^\nu =-fdt^2+\frac{dr^2}{f}+r^2(d\theta ^2+{\rm sin}^2\theta d\varphi ^2),\nonumber\\
&&f(r)=1-\frac{2M}{r}+\frac{Q^2}{r^2}\label{eq1},
\end{eqnarray}
where the  $x^\mu $ are spatial-temporal coordinates, $\mu , \nu =0, 1, 2, 3$, and $M$ and $Q$  are the mass and charge of the object, respectively.   Mikhail \textit{et al.} evaluated the non-vanishing components of the electromagnetic field tensor  $F^{\mu\nu}$. They studied effects of rotation. Taking the electric charge  to be zero, they  found an extra term in the electromagnetic tensor, corresponding to a magnetic field. Thus,  magnetic fields are generated as a result of the  rotation of the body,  as in the S-B conjecture.  The surface magnetic field for a rotating body of mass M (grams), radius R (cm),  and angular velocity $\mathbf{\Omega} ~(sec^{-1})$ is 

\begin{equation}
\mathbf{B_{p}} = \frac{9}{4}\sqrt{\frac{2M}{R}} G_{\rm N} \mathbf{\Omega} ~G. 
\label{Bp1}
\end{equation}
Comparing Eq. (\ref{Bp1}) with Eq. (\ref{m}), we have
\begin{equation}
\mathbf{B_{p}} = \frac{4\beta G_{\rm N}^{1/2}}{5Rc}M\mathbf{\Omega}~G
\label{Bp}
\end{equation}
and obtain
\begin{equation}
\beta = \frac{45c}{8G_{\rm N}^{1/2}\phi^{1/2}},
\end{equation}
where $\phi = 2M/R$. In units of  solar mass ($M_{\odot}$) and solar radius ($R_{\odot}$), we have
\begin{equation}
\beta  \approx  2730\left[\frac{R}{R_{\odot}}\frac{M_{\odot}}{M}\right]^{1/2}.
\end{equation}
For a black hole,   M/R is approximately a constant and  $\beta \sim 5$. For  a 2 solar mass neutron star,  $\beta \sim 10$.

\section{Central Engine of the Gamma-Ray Burst}

   We first  assume that the central rotating  compact object is a black hole. At the end of the section,  NMGEC-MW is applied to a rotating  proto-magnetar. 

The maximum amount of energy
which can be extracted from a  black hole is
the rotational energy, $E_{rot}$
  \begin{equation}
  E_{rot} = Mc^{2}-M_{irr}c^{2},
  \end{equation}
  where  $M_{irr}$ is the irreducible   mass of the black hole, 
   \begin{equation}
  M_{irr} = \sqrt{\frac{S_{BN}}{4\pi k_{B}}}M_{Planck}.
  \end{equation}
  $S_{BN}=A_{BH}k_{B}c^{3}/4G_{\rm N}\hbar$ is  the entropy,
    $A_{BH}$  the surface area,  and $M_{Planck} = \sqrt{c\hbar/G_{\rm N}}$ the Planck mass.
 The rotational energy of a black hole with angular momentum $J$ is a fraction of the black hole mass $M$,
 \begin{eqnarray}
 E_{rot} &=& f(\alpha)Mc^{2},\\
 f(\alpha) &=& 1- \sqrt{\frac{1}{2}[1+\sqrt{1-\alpha^{2}}]},
 \end{eqnarray}
where $\alpha = Jc/M^{2}G_{\rm N}$ is the rotation parameter. For a maximally rotating black hole $(\alpha = 1), f = 0.294$.

From Eq. (\ref{Bp}), the magnetic field at the black hole horizon is
\begin{eqnarray}
B_{p} &=& \frac{9}{4}M\sqrt{\frac{5G_{\rm N}\alpha}{R^{3}c}} \approx 8.13\times10^{8}\left(\frac{M}{M_{\odot}}\right)\alpha^{1/2}\left(\frac{R}{R_{\odot}}\right)^{-3/2} G.
\label{Bgama}
\end{eqnarray}

In order to evaluate Eq. (\ref{Bgama}), we  use characteristic parameters for the black hole:  $\alpha \sim 0.1-1$ and  M $\sim 2.5 ~  M_{\odot}$ \cite{lee00}.  The horizon radius for a rotating  Kerr  black hole is
\begin{equation}
r_{BN} = \frac{r_{Sh}}{2}\left[1+\sqrt{1-\alpha^{2}}\right],
\end{equation}
where $r_{Sh} = 2G_{\rm N}M/c^{2}$ is the Schwarzschild radius.

 The prediction for magnetic fields in  GRBs, using  Eq.  (\ref{Bgama}), is $B_{p} \sim 10^{15}-10^{16}$ G. We  conclude that the  NMGEC-MW theory predicts  the required magnetic fields in rotating black holes. 
For a rapidly rotating proto-magnetar, we obtain  from Eqs. (2.1)-(2.3) and (3.5)
\begin{equation}
B_{p} =  5.4 \times 10^{13} \beta P^{-1} ~ G, 
\label{bpfim}
\end{equation}
where $\beta \sim 10$ and P is the period in seconds. For $P\ll1 ~  s$,    we obtain $B_{p} \gtrsim 10^{15}$ G. 

\section{Conclusions and Discussion}

Magnetic fields $\simeq 10^{15}-10^{16}$ G have been suggested to be connected with central engines of GRBs. 
 We considered the possibility that GRBs are powered by a rapidly rotating highly magnetized central compact object, a black hole or a neutron star.   Since these magnetic fields are  difficult to be obtained   by  astrophysical mechanisms, we considered the possibility that the fields are created by NMGEC-MW. 

For  GRBs a magnetic field $\sim 10^{15}-10^{16}$ G is required to produce the
Poynting flux needed to supply the energy observed within the  required time.  The  fields  predicted by  NMGEC-MW in Eq.  (\ref{Bgama}) for a black hole and Eq. (\ref{bpfim}) for a neutron star,  are   in agreement with models  requiring central engines with $\gtrsim 10^{15}$ G fields. We conclude that NMGEC-MW is a possible mechanism for  creating $\gtrsim 10^{15}$ G fields in the central engine of GRBs due to a rotating neutron star or black hole.

From the results of our paper, one might believe that stellar mass black holes lose the rotation generated in their formation very quickly due to magnetic dipole emission from the strong magnetic field generated by NMGEC-MW. We might then conclude that only the youngest stellar black holes observed have non-zero rotation. However, stellar mass black holes are only observed in accreting binaries. Accretion from the stellar companion makes the black hole visible in X-rays. The accretion is able to spin-up the black holes. Thus, old black holes, as well as young black holes,  can have non-zero rotation.

\clearpage

\acknowledgments

R.S.S. thanks the Brazilian agency FAPESP for financial support
(04/05961-0). R.O. thanks FAPESP (06/56213-9) and the Brazilian
agency CNPq (300414/82-0) for partial support. We would like to thanks an unidentified referee for helpful comments.


\begin{thebibliography}{999}


\bibitem{pir05b} Piran T., \textit{AIPC} \textbf{784} (2005) 164.
\bibitem{lee00} Lee, H. K.; Wijers, R. A. M. J. and Brown, G. E., \prep{325}{2000}{83}.
\bibitem{pir05} Piran, T., \rmp{76}{2005}{1143}
\bibitem{uso92}Usov, V. V., \nature {357}{1992}{472}.
\bibitem{tho94} Thompson, C.,  \textit{Mon. Not. R. Astron. Soc.}\textbf{ 270} (1994) 480. 
\bibitem{kat97} Katz, J. I., \apj{490}{1997}{633}.
\bibitem{klu98} Klu{\'z}niak W.,  and Ruderman M., \apj{508}{1998}{113}.
\bibitem{whe00} Wheeler, J. C., Yi, I.,  H\"{o}flich, P.,  Wang, L., \apj{537}{2000}{810}. 
\bibitem{tho04} Thompson, T. A.,  Chang, P.,  Quataert, E., \apj{611}{2004}{380}
\bibitem{met07} Metzger, B.  D.,  Thompson, T. A.,  Quataert, E. \apj{659}{2007}{561}.

\bibitem{buc06} Bucciantini, N.,  Thompson, T. A.,  Arons, J.,  Quataert, E., and  Del Zanna, L., \textit{Mon. Not. R. Astron. Soc.} \textbf{368} (2006) 171.
\bibitem{kom07} Komissarov, S. S., and  Barkov, M.  V., \textit{Mon. Not. R. Astron. Soc.} \textbf{382} (2007) 1029.
\bibitem{buc07} Bucciantini, N.,  Quataert, E.,  Arons, J.,  Metzger, B. D.,  Thompson, T. A., \textit{Mon. Not. R. Astron. Soc.} \textbf{380} (2007) 1541.
\bibitem{yu07} Yu, Y. W., and  Dai, Z. G., \textit{Astron. Astrophys.} \textbf{470} (2007) 119. 
\bibitem{tho07} Thompson, T. A., \textit{Rev. Mex. Astron. Astrophys.} \textbf{27} (2007) 80
\bibitem{spr08} Spruit, H. C., \textit{AIPC} \textbf{983} (2008) 391.
\bibitem{sch11} Schuster, A., 
\textit{Proc. Phys. Soc. London.} \textbf{24}  (1911) 121. 
\bibitem{bla47} Blackett, P. M. S., \nature{159}{1947}{658}.
\bibitem{pau33}Pauli, W., \textit{Ann. Phys.} (Leipzig) \textbf{18}  (1933) 305.
\bibitem{ben49} Bennet \textit{et al.}, \textit{Proc. R. Soc. London A} \textbf{198}  (1949) 39.
\bibitem{pap50} Papapetrou, A., \textit{Philos. Mag} \textbf{41} (1950) 399.
\bibitem{luc52} Luchak, G., \textit{Can. J. Phys} \textbf{29} (1952) 470.
\bibitem {bar85} Barut, A. O. and  Gornitz, T., \textit{Found. Phys.} \textbf{15}  (1985) 433.
\bibitem{oph97} Opher, R. and Wichoski, U.
F., \prl {78} {1997} {787}.
\bibitem{mik95} Mikhail, F. I., Wanas, M. I. and  Eid, A. M., \textit{Ap$\&$SS} \textbf{228}  (1995) 221.
\bibitem{mik77} Mikhail, F. I. and Wanas, M. I., \textit{Proc. R. Soc. London. A} \textbf{356} (1977) 471.
\bibitem{wil23}  Wilson, H. A.,  \textit{Proc. R. Soc. London A} \textbf{104}
(1923) 451.
\bibitem{sir79}Sirag, S. P., \nature{278}{1979}{535}.
\bibitem{mik81} Mikhail, F. I. and Wanas, M. I.,  \textit{Int. J. Theor. Phys.} \textbf{20}(1981) 671.
\bibitem{mik85} Wanas, M. I.,  \textit{Int. J. Theor. Phys.} \textbf{24}(1985) 638.


























\end{thebibliography}
 \end{document}